\documentclass{PoS}
\usepackage{epsfig}

\title{HISQ action in dynamical simulations}

\ShortTitle{HISQ action in dynamical simulations}

\author{\speaker{A.~Bazavov}$^a$, C.~Bernard$^b$, C.~DeTar$^c$, 
        W.~Freeman$^a$, Steven~Gottlieb$^d$, U.M.~Heller$^e$, J.E.~Hetrick$^f$,
        J.~Laiho$^b$, L.~Levkova$^c$, J.~Osborn$^g$, R.~Sugar$^h$, 
        and D.~Toussaint$^a$\newline(MILC Collaboration)\\
        \llap{$^a$} Department of Physics, University of Arizona, Tucson, AZ 85721, USA\\
        \llap{$^b$} Department of Physics, Washington University, St.~Louis, MO 63130, USA\\
        \llap{$^c$} Physics Department, University of Utah, Salt Lake City, UT 84112, USA\\
        \llap{$^d$} Department of Physics, Indiana University, Bloomington, IN 47405, USA\\
        \llap{$^e$} American Physical Society, One Research Road, Box 9000, Ridge, NY 11961, USA\\
        \llap{$^f$} Physics Department, University of the Pacific, Stockton, CA 95211, USA\\
        \llap{$^g$} Argonne Leadership Computing Facility, Argonne National Laboratory, \\Argonne, IL 60439, USA\\
        \llap{$^h$} Department of Physics, University of California, Santa Barbara, CA 93106, USA
}

\abstract{
We report on recent progress in employing the 
Highly Improved Staggered Quark (HISQ) action
introduced by the HPQCD/UKQCD collaboration in simulations 
with dynamical fermions. The HISQ action is an order $a^2$ Symanzik-improved 
action with further suppressed taste symmetry violations.
The improvement in taste symmetry is achieved by introducing Fat7 smearing 
of the original gauge links and reunitarization (projection to an element 
of U(3) or SU(3)) followed by Asq-type smearing.
Major challenges for calculating the fermion force are related 
to the reunitarization step. We present a preliminary study 
of the HISQ action on two 2+1+1 flavor ensembles with the lattice spacing 
roughly equivalent to the MILC asqtad $a=0.125$ and 0.09 fm ensembles.
}

\FullConference{The XXVI International Symposium on Lattice Field Theory \\
                 July 14 - 19, 2008\\
                 Williamsburg, Virginia, USA}

\begin{document}

\section{Introduction}

The Highly Improved Staggered Quark (HISQ) action developed in 
Ref.~\cite{HPQCD07} is an $O(a^2)$ Symanzik-improved action 
for which additional suppression of 
taste-exchange interactions is \newline achieved by replacing the original gauge
links $U$ in the Dirac operator by
\begin{equation}
  U\rightarrow X={\cal F}_2{\cal U}{\cal F}_1\,U
\end{equation}
where intermediate sets of links $V$, $W$, $X$ are defined as
\begin{itemize}
    \item ${\cal F}_1$ -- smearing level 1 (Fat 7): $V={\cal F}_1U$,
    \item ${\cal U}$ -- reunitarization: $W={\cal U}V$,
    \item ${\cal F}_2$ -- smearing level 2 (Asq): $X={\cal F}_2W$.
\end{itemize}

A new feature of the HISQ action compared with asqtad is the reunitarization step that
gives an extra contribution to the fermion force.

\section{Fermion force and reunitarization}

In molecular dynamics simulations we are sampling an ensemble of gauge
configurations weighted by $\exp(-S)$, where the action $S=S_g+S_f$
is split into gauge, $S_g$, and fermionic, $S_f$, parts. As usual,
the integration over Grassmann variables is performed and then
a set of pseudo-fermion fields $\Phi$ is introduced, resulting in
the fermionic part of the form 
$S_f\sim \langle\Phi|(M^\dagger(U)M(U))^{-N_f/4}|\Phi\rangle$.
One can consult the details of the algorithm in \cite{MD87a} and 
recent ideas on efficient evaluation of the fermion force
for HISQ in \cite{WWlat07}.

The fermion force is calculated by taking the derivative of the action
$S_f$ with respect to fundamental gauge links $U$ using the chain
rule along the lines of Refs.~\cite{WWlat07}, \cite{FLIC04}. 
Schematically:
\begin{equation}
  \frac{\partial S_f}{\partial U}=
  \frac{\partial S_f}{\partial X}\,\frac{\partial X}{\partial W}\,
  \frac{\partial W}{\partial V}\,\frac{\partial V}{\partial U}\ .
\end{equation}
The following parts are the same as for the asqtad action:
$\partial S_f/\partial X$, $\partial X/\partial W$, 
$\partial V/\partial U$,
while the contribution from the 
reunitarization step $\partial W/\partial V$ is new.

We have experimented with projecting to SU(3) and U(3) groups
and found that pion spectrum measurements with valence HISQ
and sea asqtad quarks show no difference, as one would expect
on physical grounds. Therefore in our dynamical HISQ simulations
we always perform projection to the group U(3). While keeping the
physics the same, this has two advantages:
\begin{enumerate}
  \item Different methods of projection, namely polar decomposition and
  trace maximization, give identical results. (The same is not true
  in the SU(3) case.)
  \item When projecting from U(3) to SU(3) by making the determinant of the
  matrix equal to 1, one needs to choose among
  three possibilities for the phase. In dynamical simulations one 
  needs to track this phase to make sure it changes smoothly for each
  link. Otherwise discontinuous changes lead to rapid changes in the
  action similar to those we describe below.
\end{enumerate}

We adopt the following method of U(3) projection. 
For a complex matrix $V$ the matrix $H=V^\dagger V$
is Hermitian and $W=VH^{-1/2}$ is unitary. We first calculate
$H^{-1/2}$ by using the Cayley-Hamilton theorem in a manner
similar to the approach in
Refs.~\cite{MP04}, \cite{Hasenfratz:2007rf}:
\begin{equation}\label{H12}
  H^{-1/2}=f_0\mathbb{I}+f_1H+f_2H^2,
\end{equation}
where $f_i$ are functions of $c_i={\rm Tr}(H^{i+1})/(i+1)$,
$i=0,1,2$. To evaluate the derivative of $H^{-1/2}$ with respect to
$H$ one needs to know the derivative of $f_i$ with respect
to $H$. This can be performed analytically by applying the chain
rule and making use of the coefficients explicitly calculated in
Ref.~\cite{Hasenfratz:2007rf}. After $\partial H^{-1/2}/\partial H$
is known we have trivially:
\begin{equation}\label{dWdV}
  \frac{\partial W}{\partial V}=
  \frac{\partial V}{\partial V}H^{-1/2}+
  V\frac{\partial H^{-1/2}}{\partial H}\frac{\partial H}{\partial V}.
\end{equation}
(We write Eq.~(\ref{dWdV}) schematically, but in fact each matrix-matrix
derivative is a rank 4 tensor with indices contracted in such a way
that the resulting expression on the right hand side is a rank 4 tensor again.
An excellent review on how to deal with such derivatives can be found
in \cite{FLIC04}.) Thus the entire derivation can be performed analytically. 
(We also implemented a finite difference scheme and rational function 
approximation but found them much less accurate than the 
procedure described here.)

To understand the possible behaviour of
$\partial W/\partial V$ qualitatively let us consider the U(1) case
for the moment. Then $V=r\,e^{i\theta}$ and
$W=V(V^\dagger V)^{-1/2}=e^{i\theta}$, and the derivative
\begin{equation}\label{derU1}
  \frac{\partial W}{\partial V} =
  \left(\frac{\partial W}{\partial V}\right)_{V^\dagger}
  =\frac{\partial(W,V^\dagger)}{\partial(V,V^\dagger)}=
  \frac{\partial(W,V^\dagger)}{\partial(r,\theta)}\,
  \frac{\partial(r,\theta)}{\partial(V,V^\dagger)}=\frac{1}{2r}
\end{equation}
is large when $r$ is small, as might happen when the links $U$ are locally
disordered, and the first stage of smearing results in a small smeared
link $V$. For the matrix case the derivative 
is dominated by the smallest eigenvalue of $V$. To derive Eq.~(\ref{derU1})
we applied the method of Jacobians and the notation is similar to the 
one used in Thermodynamics.

\section{Dynamical HISQ simulations}

We run the RHMC algorithm with five pseudofermion fields.  The first
pseudofermion implements the ratio of the determinants for two
light and one strange quark to the determinant for three unphysical
heavy quarks (``UHQ'') with mass $am_{UHQ}=0.2$.  The next three
each implement the determinant for one UHQ, and the final pseudofermion
implements a physical charm quark, including the mass correction to the
Naik term to first order in $m$.

At the time we did our studies, one loop fermion corrections 
to the gauge action were not yet known, so we used 
the coefficients in our gauge action appropriate
for the asqtad fermion action. (The coefficients of the gauge action
with 1-loop corrections due to HISQ have been calculated and are now
available \cite{HHHPoS08}.)

Our integration algorithm is the ``3G1F'' (``three gauge steps, one fermion step'')
algorithm, with the Omelyan integrator used for both gauge and fermion forces.
Our convention for the step size is that each application of the fermion
force is one step.  Note that since the Omelyan integrator shifts the
time at which the force is calculated alternately forward and backward,
the full period in simulation time of the integrator is twice what we
call the step size.  For example, when we run a fifty-step trajectory
this means 25 cycles of the Omelyan integrator, involving 50 fermion
force calculations and 150 gauge force calculations. The parameters
of the three 2+1+1 flavor ensembles we have run are compiled in 
Table~\ref{tab_3ens}. The lattice spacing is calculated by measuring
the ratio $r_1/a$ from the static quark potential 
and using the value $r_1=0.318$ fm \cite{MILCspec2004}.


\begin{table}
\centering
\begin{tabular}{|l|r|l|l|l|l|l|l|}   \hline
$N_s^3\times N_t$ & Conf. & $\beta$ & $am_l$ & $am_s$ & $am_c$ & 
  $\Delta t$ & $a$, fm\\\hline
$20^3\times64$ & 40 & 6.75 & 0.010 & 0.050 & 0.600 & 0.04167 & 0.127    \\
$28^3\times96$ & 50 & 7.07 & 0.007 & 0.035 & 0.420 & 0.03125 & 0.093    \\
$48^3\times144$&  6 & 7.47 & 0.004 & 0.020 & 0.240 & 0.01250 & 0.060    \\
\hline
\end{tabular}
\caption{Dynamical HISQ 2+1+1 ensembles.}\label{tab_3ens}
\end{table}

\begin{figure}
\begin{center}
\epsfig{figure=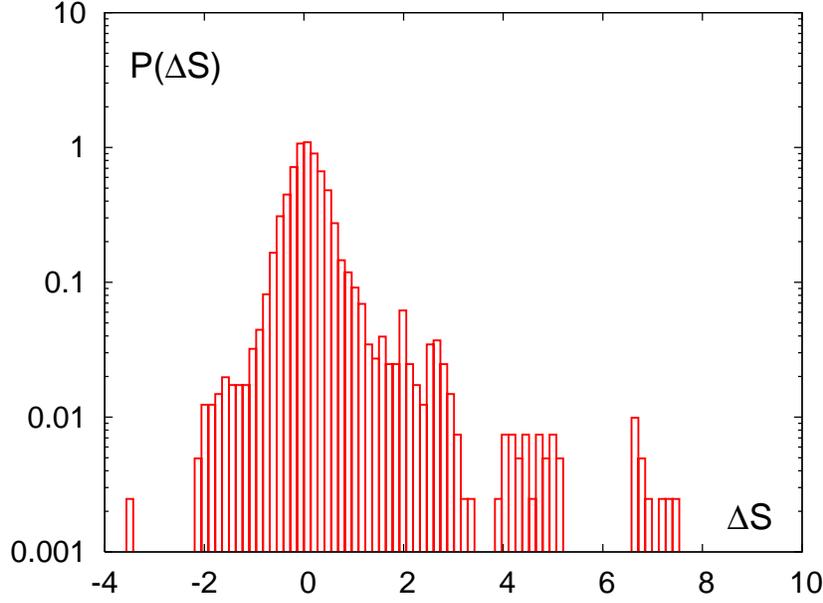,width=0.8\textwidth}
\caption{The histogram of the change in the action.}
\label{fig_delta_hist}
\end{center}
\end{figure}

While most trajectories ran smoothly, we found many trajectories with
large jumps in the action as the integration proceeded.
In Fig.~\ref{fig_delta_hist} we show the histogram
of the change in the action, plotted on a logarithmic scale,
over several time units for the $a=0.127$ fm
ensemble. The long tail ``outliers'' indicate instantaneous jumps 
in the action, which we investigate further.

\begin{figure}
\begin{center} 
\epsfig{figure=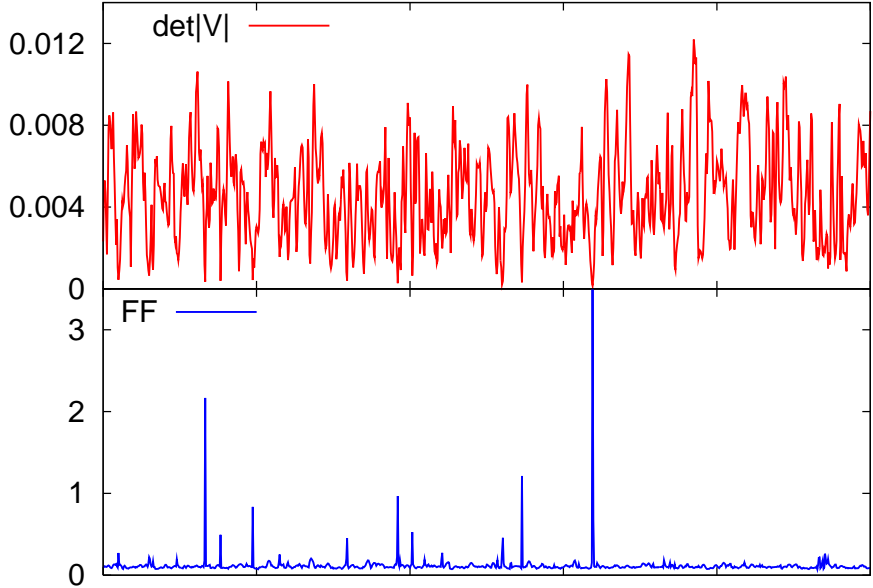,width=0.8\textwidth}
\caption{The time history of $|\det V|_{min}$ (top) and 
$||F||_{max}$ (bottom) 
for the $a=0.127$ fm ensemble during 5 time units ($=5\times24=120$ time
steps).}
\label{fig_force_detV}
\end{center}
\end{figure}

Let us denote the norm of a matrix $A$ by:
$$
    ||A||=\sqrt{\sum_{i,j}|A_{ij}|^2}.
$$
When we calculate the fermion force at each time step as an anti-Hermitian
matrix defined on each link, we evaluate its norm and find the maximum
value over the lattice: $||F||_{max}$. Also, at each time step
we calculate the determinant of $V$ (Fat7 smeared) links and
find the minimum value over the lattice: $|\det V|_{min}$.
Time histories of these two quantities are shown in 
Fig.~\ref{fig_force_detV}. Large values of the fermion force
accompany small values of the determinant (or small eigenvalues) 
of Fat7 smeared links. Thus, when calculating smeared
links by adding different paths one may by chance produce a matrix
$V$ with an eigenvalue close to 0, which in turn leads to a large
derivative $dW/dV$ (as in the U(1) example considered earlier) that
results in a large fermion force. Our integration algorithm has a finite
step size, so it is not able to integrate such ``spikes'' in the force smoothly,
leading to instantaneous jumps in the action that we see as 
``outliers'' in the action histogram Fig.~\ref{fig_delta_hist}.

\begin{figure}
\begin{center}
\epsfig{figure=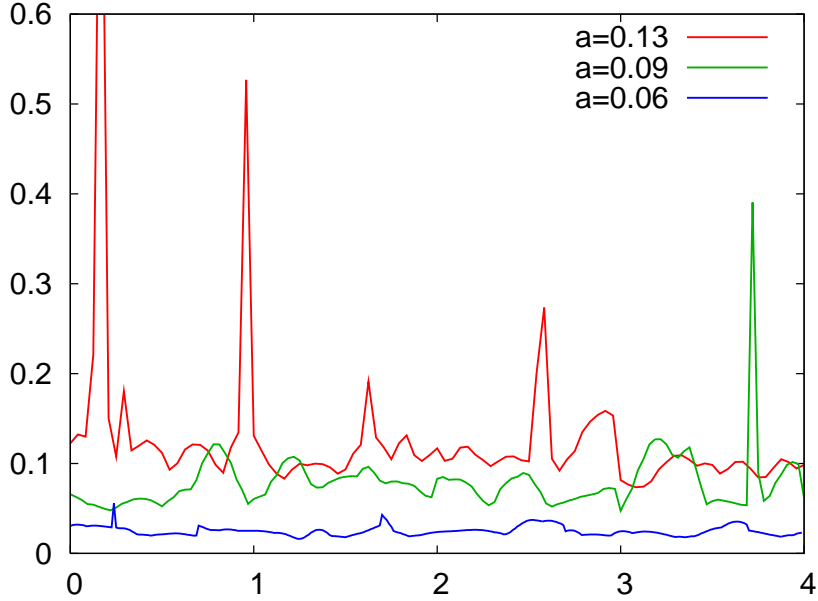,width=0.8\textwidth}
\caption{Fermion force $||F||_{max}$ for different ensembles
versus time in time units (1 time unit corresponds to 24 steps for
$a=0.127$, 32 steps for $a=0.093$ and 80 steps 
for $a=0.060$ ensemble).}
\label{fig_force_3ens}
\end{center}
\end{figure}

We investigated how this situation changes when we go to finer
(smaller lattice spacing) ensembles. Since configurations become
smoother, spikes in the force become less severe, as seen in
Fig.~\ref{fig_force_3ens}.

\section{Pion splittings on dynamical HISQ configurations}

The effect of suppression of the taste-exchange interactions in HISQ
was investigated in \cite{HPQCD07} by measuring the pion spectrum
with valence HISQ on sea asqtad configurations. Here we report
on similar measurements performed on dynamical HISQ configurations
for the first two ensembles of Table~\ref{tab_3ens}.

It is convenient to define a dimensionless 
quantity which is almost independent of quark mass:
\begin{equation}
  \Delta\equiv (M^2_\pi-M^2_{G})r_1^2,
\end{equation}
where $M_G$ corresponds to the Goldstone pion and
$M_\pi$ refers to one of the other seven pion
tastes in Tables~\ref{tab_a127} and \ref{tab_a093}.
We calculate $\Delta$ for comparable asqtad and HISQ configurations,
and then the ratio
\begin{equation}
R\equiv \frac{\Delta_{ASQ}}{\Delta_{HISQ}}
\end{equation}
shows how much the splittings decrease when going from asqtad to HISQ.
The values of $R$ for different pion tastes are shown in the last column
of Tables~\ref{tab_a127} and \ref{tab_a093}. The statistical errors
are rather large since for HISQ we have about an order of magnitude fewer
configurations than for asqtad. The number of configurations used for
measurements is indicated in parentheses in the headers of the second and
third column. The overall trend is however clear and in agreement
with Ref.~\cite{HPQCD07}: about a factor of three improvement in taste symmetry
for the HISQ action relative to asqtad.

\begin{table}
\centering
\begin{tabular}{|c||l|l||l|l||l|}   \hline
Pion taste & $r_1\,M^{ASQ}_\pi$(658)  & $r_1\,M^{HISQ}_\pi$(40) & 
    $\Delta_{ASQ}$ & $\Delta_{HISQ}$ & $R$ \\ \hline
$\gamma_5$         & 0.2244(02) & 0.1889(07) &          &           &     \\
$\gamma_0\gamma_5$ & 0.2815(11) & 0.2071(27) & 0.029(1) & 0.0072(11)& 4.0(6) \\
$\gamma_i\gamma_5$ & 0.2822(05) & 0.2058(10) & 0.029(0) & 0.0067(05)& 4.4(3) \\
$\gamma_i\gamma_j$ & 0.3134(20) & 0.2224(33) & 0.048(1) & 0.0138(15)& 3.5(4) \\
$\gamma_i\gamma_0$ & 0.3126(11) & 0.2188(19) & 0.047(1) & 0.0122(08)& 3.9(3) \\
$\gamma_i$         & 0.3347(28) & 0.2306(56) & 0.062(2) & 0.0175(26)& 3.5(5) \\
$\gamma_0$         & 0.3373(15) & 0.2311(22) & 0.063(1) & 0.0178(10)& 3.6(2) \\
$\mathbb{I}$       & 0.359(5)   & 0.252(12)  & 0.079(4) & 0.0280(61)& 2.8(6) \\
\hline
\end{tabular}
\caption{Pion spectrum on $a=0.127$ fm HISQ ensemble.}\label{tab_a127}
\end{table}

\begin{table}
\centering
\begin{tabular}{|c||l|l||l|l||l|}   \hline
Pion taste & $r_1\,M^{ASQ}_\pi$(572)  & $r_1\,M^{HISQ}_\pi$(50) & 
    $\Delta_{ASQ}$ & $\Delta_{HISQ}$ & $R$ \\ \hline
$\gamma_5$         & 0.2069(05) & 0.1378(08) &          &           &     \\
$\gamma_0\gamma_5$ & 0.2177(10) & 0.1420(08) & 0.0046(5)& 0.0012(4) & 4(1)   \\
$\gamma_i\gamma_5$ & 0.2187(07) & 0.1428(08) & 0.0050(4)& 0.0014(3) & 3.6(8) \\
$\gamma_i\gamma_j$ & 0.2256(11) & 0.1467(21) & 0.0081(5)& 0.0025(7) & 3.2(8) \\
$\gamma_i\gamma_0$ & 0.2259(07) & 0.1475(11) & 0.0082(4)& 0.0028(4) & 3.0(4) \\
$\gamma_i$         & 0.2311(15) & 0.1485(16) & 0.0106(7)& 0.0031(5) & 3.5(6) \\
$\gamma_0$         & 0.2318(10) & 0.1509(11) & 0.0109(5)& 0.0038(4) & 2.9(3) \\
$\mathbb{I}$       & 0.2398(25) & 0.1522(27) & 0.015(1) & 0.0042(9) & 3.5(8) \\
\hline
\end{tabular}
\caption{Pion spectrum on $a=0.093$ fm HISQ ensemble.}\label{tab_a093}
\end{table}

\section{Conclusions}

In dynamical HISQ simulations with typical parameters, we found
that smearing may produce a smeared link $V$ with a small eigenvalue that dominates
the derivative of the reunitarized link $W$, $dW/dV$, and gives a large
contribution to the fermion force. Such ``spikes'' in the force integrated
with finite time steps give ``jumps'' in the action that decrease the
acceptance rate of the RHMC algorithm. This problem was noted in
Ref.~\cite{Hasenfratz:2007rf} and is probably related to topological
defects, ``dislocations'' that manifest themselves in plaquettes with
low values lying on the tails of the plaquette distribution.
We found that going to finer ensembles (smoother gauge configurations)
reduces the number of spikes and partially cures the problem.

On configurations generated with the HISQ action for dynamical quarks
we measured the staggered pion spectrum and 
found that pion splittings decrease by a factor of three or more, confirming
the result of Ref.~\cite{HPQCD07}, where pioneering tests were done
with valence HISQ fermions on configurations with asqtad sea quarks.

\end{document}